# A Systematic Analytical Design Procedure for Distributed Amplifiers

Elham AMIRI† *and* Mojtaba JOODAKI††

**SUMMARY** In this paper we present a simple while comprehensive analytical design procedure for distributed amplifiers. Distributed amplifiers are attractive for designers due to their wideband capability. When designing a distributed amplifier, the first question that comes to mind is how wide the bandwidth can be. This paper answers this question by using the self-matching and low-pass properties of a distributed amplifier. Self-matching property of a distributed power amplifier is an interesting point that distinguishes it from other types of power amplifiers that are usually based on input and output matching networks. Here the estimation of the bandwidth of a distributed amplifier structure is discussed. The equations that are used in this paper can bring good insight and they can assist designers. Furthermore, we have explained the frequency behavior of a tapered distributed amplifier analytically for the first time. In order to validate the approach presented here, we have used published designs including our previously published design as practical examples. The flowchart of the design procedure is also provided.
*key words: distributed amplifier, non-uniform distributed amplifier, power amplifier, transmission line, wideband amplifier.*

## 1. Introduction

The need for higher data rates is increasing day by day. Therefore, paying attention to wideband amplifiers is inevitable. Distributed amplifier is one of the most interesting ideas for a wideband design. In a distributed amplifier (DA) configuration several parallel transistors are connected to each other between two transmission lines (TL), as can be seen in Fig. 1. Despite the fact that DA is not capable of high gain, it provides a good matching over a wide bandwidth. Ideally, a DA engages parasitic capacitances of active cells into input and output lines, forming two new artificial TLs. Consequently, the capacitances will not limit the bandwidth anymore.

If a DA designer is interested in a frequency range, they need to know how to choose a transistor or if a certain transistor can be used in a specific design or not. In this paper, we use the parameters of the active cells (i.e. transistors) to predict the maximum attainable frequency range of a DA. When designing a DA, without knowing the expected bandwidth, we cannot be sure if our design is near to the optimum solution. And also, during the optimization process, even using simulations, the designer may need to know the bandwidth limit in beforehand to prevent wasting time on blindly sweeping different parameters that may not improve the bandwidth [1]. In this work, by presenting a design procedure for a standard or conventional DA, we will avoid these uncertainties. This approach can be used as the first stage of the design and it is applicable to any type of DA including uniform and non-uniform DAs.

The section 2 of this paper presents the design procedure and the maximum achievable bandwidth of a DA. The discussion about this procedure will be covered in the section 3. Section 4 talks about the inductive parts that should be implemented. The power performance issues of the design are included in section 5. In this paper, we consider the conventional design and also two widespread unconventional techniques known as tapering and gate series capacitors methods.

## 2. Frequency Behavior of the Design

When designing a single transistor amplifier, by increasing the operating frequency the impedances of parasitic capacitances reduce gradually. After a frequency point, these impedances are so low that the main signal is shorted and in fact no signal reaches the output. If the parasitic capacitances of a transistor are not controlled, they will limit the bandwidth tremendously. DA structure is a technique to control and absorb these troublesome parasitic capacitances. A DA accomplishes this duty by resembling an artificial TL. This trick nullifies the demeaning effect of the parasitic capacitances. Fig. 1 shows two artificial TLs that are built at the gates and drains of the transistors in a DA configuration. These new artificial TLs are constructed of repeating patterns of inductors and parasitic capacitances. Each TL is an infinite ladder network of T-section constant-k filters. The T-models of these TLs are presented in Fig. 2. The inductance and capacitance of a conventional TL are usually reported per unit length but in an artificial TL, it is preferred to be given per unit cell. The main parameter of a TL is its characteristic impedance ($Z_0$) [2]. Considering lossless conditions, the characteristic impedances of the artificial TLs in the gate and drain are given by:

$$z_{0(G)} = \sqrt{\frac{L_{(G)}}{C_{(G)}}} \qquad (1)$$

$$z_{0(D)} = \sqrt{\frac{L_{(D)}}{C_{(D)}}} \qquad (2)$$

---

†The author is with Department of Electrical Engineering, Ferdowsi University of Mashhad, Mashhad, Iran.
††The author is with the School of Computer Science and Engineering, Constructor University, Bremen, Germany, (e-mail: mjoodaki@constructor.university).

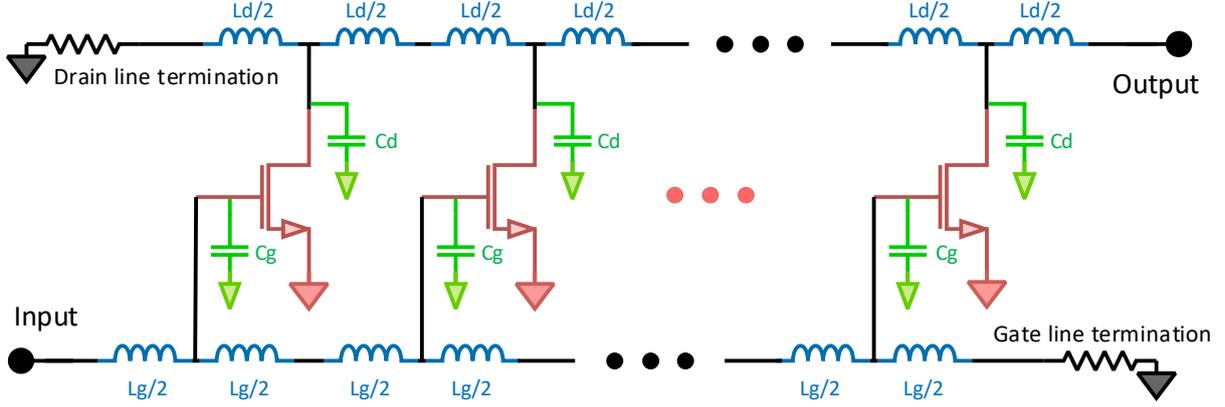

**Fig. 1** Schematic of a conventional distributed amplifier (DA) circuit.

where $L_{(G)}$ and $L_{(D)}$ are the inductances and $C_{(G)}$ and $C_{(D)}$ are the capacitances of each cell in these artificial TLs. The impedance of the input source and the output load that a DA deals with are typically 50 $\Omega$. So the two artificial TLs are designed to have 50 $\Omega$ characteristic impedances. The advantage of a DA comes from this self-matching nature. The bandwidth of a DA is enhanced greatly because both the input and the output will be matched to 50 $\Omega$ up to the cut-off frequency of the transmission lines. Thus, the bandwidth of the design is limited to the cut-off frequency of these artificial transmission lines. This means a DA has a low-pass nature. Therefore, we focus on this clue to carry on the design procedure. The cut-off frequency of a TL is given by [3]:

$$f_{cut-off\,(TL)} = \frac{1}{\pi\sqrt{L_{(TL)}C_{(TL)}}} \tag{3}$$

Considering (3), the gate line and drain line cut-off frequencies are as follows, respectively:

$$f_{cut-off\,(G)} = \frac{1}{\pi\sqrt{L_{(G)}C_{(G)}}} = \frac{1}{\pi Z_{0(G)}C_{(G)}} \tag{4}$$

$$f_{cut-off\,(D)} = \frac{1}{\pi\sqrt{L_{(D)}C_{(D)}}} = \frac{1}{\pi Z_{0(D)}C_{(D)}} \tag{5}$$

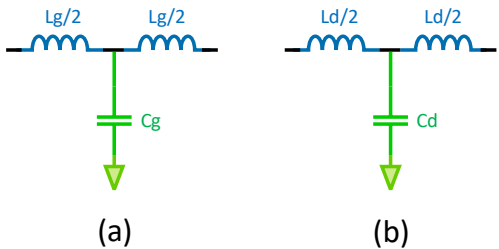

**Fig. 2** a) Gate line T-model. b) Drain line T-model.

The biggest parasitic capacitance of the active cell (i.e. the transistor) is the bottle neck of a DA design. That is generally the gate-source capacitance of the transistor. For instance, the gate-source capacitance of TGF2023-01 transistor from Qorvo is 1.79 pF that is six times bigger than the drain-source capacitance of the transistor. Therefore, the $C_{GS}$ is the capacitance in (1) and (4). This dominating capacitance is repeatedly absorbed in the artificial transmission line that is connected to the gates of the transistors. We assume the other parasitic capacitances are negligible. As the capacitance of gate is bigger than that of drain, the cut-off frequency of the gate artificial transmission line $f_{cut-off(G)}$ is smaller than that of drain. This cut-off frequency is the upper limit of the amplifier's bandwidth. It is worth noting that in a DA, this is the maximum achievable bandwidth that can be reached when designing a conventional DA using a specific transistor.

## 3. Some Deviations from the Conventional Design

In a DA, as the input signal travels along the gate line, the transistors get stimulated one after another. The output of the first active cell travels on the drain line until it reaches the second stage output, where they get added to each other, then the resultant signal goes to the third stage. This progression continues until the signal touches the output port. In order to have a constructive superposition of the output signals, they must be in-phase. This requires the same phase velocities for the gate and drain lines. If the signal travels a wavelength ($\lambda$) distance in a period of T, then the phase velocity of the transmission line is given by [2]:

$$\upsilon_p = \frac{\lambda}{T} = \frac{1}{\sqrt{L_{(TL)}C_{(TL)}}} \tag{6}$$

Considering (6), $L_{(G)}C_{(G)}$ and $L_{(D)}C_{(D)}$ should be the same for a similar phase velocity of the gate and the drain lines. This is usually not happening in a conventional design of DA. One can add some elements to the DA structure to achieve equal $L_{(G)}C_{(G)}$ and $L_{(D)}C_{(D)}$. An interesting technique is to put a capacitor in series with the gate-source capacitance of each transistor [4-7]. These capacitors can be chosen in order to reduce $C_{(G)}$ and make it equal to $C_{(D)}$. The equivalent capacitor value can be reduced remarkably and as a result the cut-off frequency of the amplifier might increase considerably. This idea can be applied to uniform and non-uniform DAs. The smaller the series capacitor, the higher the cut-off frequency. However, still there is a trade-off here. We cannot choose a very small capacitor, as this will reduce the transistors gain. The overall gain of DA will be reduced because a smaller portion of the input signal enters each transistor. Therefore, our ability to increase the cut-off frequency is limited.

## 4. Circuit Design Parameters

The next step of the design procedure is to determine the inductance per unit cell that provide us a 50-Ω artificial transmission line.

$$L_{(G)} = Z_{0(G)}^2 \times C_{(G)} = (50)^2 \times C_{(G)} \tag{7}$$

$$L_{(D)} = Z_{0(D)}^2 \times C_{(D)} = (50)^2 \times C_{(D)} \tag{8}$$

If we equalize the gate and drain lines capacitances, e.g. using the method mentioned above, $L_{(G)}$ and $L_{(D)}$ will be equal. These inductances can be realized by using lumped inductors or can be substituted by microstrip lines. A piece of microstrip line is illustrated in Fig. 3. The metal layer thickness, the dielectric constant and height of the substrate are t, $\varepsilon_r$ and H, respectively. The characteristic impedance of a microstrip structure is a function of its width, and its inductance is mainly depending on its length. Equation (9) explains how $Z_0$ is related to t, $\varepsilon_r$, H and the line width (W) [8]. Aiming for $Z_0$=50 Ω, the width of the line can be obtained using (10).

$$Z_0 = \frac{87}{\sqrt{\varepsilon_r + 1.41}} \ln\left(\frac{5.98H}{0.8W + t}\right) \tag{9}$$

$$W = \left(\left(7.475H \times e^{\left(\frac{Z_0\sqrt{\varepsilon_r + 1.41}}{87}\right)}\right) - 1.25t\right) \tag{10}$$

The inductance per unit length (cm) of a microstrip line ($L_m$) is correlated to its dimensions as follows [8]:

$$L_m(nH) = 2 \times \ln\left(\frac{5.98H}{0.8W + t}\right) \tag{11}$$

By putting (10) in (11):

$$L_m(nH) = \frac{2Z_0\sqrt{\varepsilon_r + 1.41}}{87} \tag{12}$$

By considering equations (1) and (2):

$$C_m(pF) = \frac{L_m}{Z_0^2} = \frac{0.264(\varepsilon_r + 1.41)}{\ln\left(\frac{5.98H}{0.8W + t}\right)}$$

$$= \frac{23\sqrt{\varepsilon_r + 1.41}}{Z_0} \tag{13}$$

Finally, the lengths of the microstrip lines which give us the required inductances per unit cell will be given by:

$$l_{(G)} = \frac{L_{(G)}}{L_{m(G)}} \tag{14}$$

$$l_{(D)} = \frac{L_{(D)}}{L_{m(D)}} \tag{15}$$

where $l_{(G)}$ and $l_{(D)}$ are the lengths of the microstrip lines that connect the gate and drain terminals of each transistor to the corresponding terminals of its adjacent active cells, respectively. By passing through these lines, the signal experiences a phase shift, which is related to the electrical length as follows:

$$\Delta\theta = 2\pi \times \frac{l}{\lambda} \tag{16}$$

If dealing with equal $C_{(G)}$ and $C_{(D)}$, these shifts should be the same in order to have identical phase velocities in the gate and drain lines. Moreover, these microstrip pieces have parasitic capacitances that are assumed insignificant in comparison with transistors parasitic capacitances.

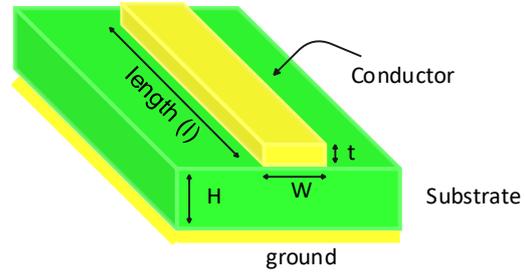

**Fig. 3** The parameters of microstrip lines.

## 5. Power Gain and Efficiency of the Design

The last step to complete the design is finding the required number of transistors (n) for a desired power gain. Assuming that the phase velocity of the gate and drain lines are equal, in a DA, the transistors outputs will be added in-phase. Therefore, if we think of ideal and lossless conditions the more stages, the higher power and voltage gains. Since the drain of each transistor sees $Z_0$ in both forward and reverse directions, the equal impedance is $Z_0/2$ and in fact, half of the current from each stage reaches the output port.

$$VoltageGain = A_V = \frac{V_{out}}{V_{in}} = \frac{I_{out} \times \frac{Z_{0(D)}}{2}}{V_{in}}$$

$$= \frac{(ng_m V_{in}) \times \frac{Z_{0(D)}}{2}}{V_{in}} = g_m \times \frac{Z_{0(D)}}{2} \times n \tag{17}$$

$$PowerGain = G_P = \frac{P_{out}}{P_{in}} = \frac{\left(|I_{out}|^2 \times \frac{Z_{0(D)}}{2}\right)}{\left(|V_{in}|^2 \times \frac{2}{Z_{0(G)}}\right)}$$

$$= g_m^2 \times \frac{Z_{0(G)} Z_{0(D)}}{4} \times n^2 \tag{18}$$

However, in practical cases, as we proceed, due to the gate line losses, a smaller input signal reaches the next transistor.

Furthermore, output signals of transistors, especially that of the first transistor in Fig. 1, would have to pass a long path on the drain line to reach the output. A longer path means a higher loss and as a result output signals will be attenuated. So, there must be an optimum number of transistors for the maximum $G_P$. In order to find this optimum number, we need to consider the attenuating behavior of a TL in a DA. Propagation constant of a TL is defined as [2]:

$$\gamma = \alpha + j\beta = \sqrt{ZY} \qquad (19)$$

where α is attenuation constant, β is phase constant, Z is the series impedance of the line per unit length and Y is the shunt admittance of the line per unit length. For gate line:

$$Z_{Gate} = R_{(G)} + j\omega L_{(G)} = j\omega L_{(G)} \qquad (20)$$

$$Y_{Gate} = G_{(G)} + j\omega C_{(G)} + \frac{j\omega C_{GS}/l_{(G)}}{1+j\omega R_i C_{GS}} = $$
$$j\omega C_{(G)} + \frac{j\omega C_{GS}/l_{(G)}}{1+j\omega R_i C_{GS}} \qquad (21)$$

At our frequencies of interest, R<<ωL and G<<ωC and can be neglected. Again for drain line:

$$Z_{Drain} = R_{(D)} + j\omega L_{(D)} = j\omega L_{(D)} \qquad (22)$$

$$Y_{Drain} = G_{(G)} + \frac{1}{R_{DS}l_{(D)}} + j\omega\left(C_{(D)} + \frac{C_{DS}}{l_{(D)}}\right) = $$
$$\frac{1}{R_{DS}l_{(D)}} + j\omega\left(C_{(D)} + \frac{C_{DS}}{l_{(D)}}\right) \qquad (23)$$

The real part of propagation constant is attenuation constant. And that is the sum of conductor loss ($\alpha_c$), dielectric loss ($\alpha_d$), radiation loss ($\alpha_r$), and leakage loss ($\alpha_l$).

$$\alpha_{(G)} = \alpha_{c(G)} + \alpha_{d(G)} + \alpha_{r(G)} + \alpha_{l(G)} = \mathrm{Re}\left(\gamma_{Gate}\right) = $$
$$\mathrm{Re}\left(\sqrt{Z_{Gate}Y_{Gate}}\right) = \frac{\omega^2 R_i C_{GS}^2 Z_0}{2l_{(G)}} \qquad (24)$$

$$\alpha_{(D)} = \alpha_{c(D)} + \alpha_{d(D)} + \alpha_{r(D)} + \alpha_{l(D)} = \mathrm{Re}\left(\gamma_{Drain}\right) = $$
$$\mathrm{Re}\left(\sqrt{Z_{Drain}Y_{Drain}}\right) = \frac{Z_0}{2R_{DS}l_{(D)}} \qquad (25)$$

When we consider losses, some parameters of (18) will change into functions of $\alpha_{(G)}$ and $\alpha_{(D)}$, i.e. $I_{out}$, $V_{out}$ and $V_{in}$. In an effort to maintain conciseness, those functions can be found in [2]. This matter will modify the power gain in (18) to:

$$PowerGain = G_P = g_m^2 \times \frac{Z_{0(G)}Z_{0(D)}}{4} $$
$$\times \left(\frac{e^{-n\alpha_{(G)}l_{(G)}} - e^{-n\alpha_{(D)}l_{(D)}}}{e^{-\alpha_{(G)}l_{(G)}} - e^{-\alpha_{(D)}l_{(D)}}}\right)^2 \qquad (26)$$

The maximum $G_P$ happens when n=$n_{opt}$. If we set the derivative of (26) equal to zero, the $n_{opt}$ will be found as [2]:

$$n_{opt} = \frac{\ln\left(\alpha_{(G)}l_{(G)}\big/\alpha_{(D)}l_{(D)}\right)}{\alpha_{(G)}l_{(G)} - \alpha_{(D)}l_{(D)}} \qquad (27)$$

Substituting (24) and (25) in (27) will result in the following equation:

$$n_{opt} = \frac{2R_{DS}\left(\ln\left(\omega^2 R_i C_{GS}^2 R_{DS}\right)\right)}{Z_0\left(\omega^2 R_i C_{GS}^2 R_{DS} - 1\right)} \qquad (28)$$

Where the $n_{opt}$ is the optimum number of transistors in Fig. 1. These transistors will get connected to one another using inductors of each cell, i.e. $L_{(G)}$ and $L_{(D)}$ in gate and drain lines, respectively. In practice, the optimum number of transistors will be significantly less than what is anticipates in (28). As n increases the difference between $G_P(n)$ and $G_P(n+1)$ decreases. This difference will become so negligible in higher number of transistors that the $G_P$ increment does not justify the costs and area increases. Therefore, designers usually choose the n between 3-6. To this point, the uniform DA design is completed.

On the other hand, despite the fact that we have a good matching in a DA configuration, the transistors will not see their optimum loads. By the term optimum load, we mean the load that leads to the highest output power or the highest efficiency. And its value often gets extracted from the load-pull figure of each transistor. Also, as already mentioned, half of the signal will go in backward direction and will get wasted in the termination resistance. Therefore, in a DA, the output power is not so high and the efficiency of the amplifier is limited. This means that the wideband performance of DAs comes at a price. Consequently, in practice, non-uniform and unconventional DA designs are usually implemented [9-12]. A common way of increasing efficiency of a DA is the tapering technique. This concept is shown in Fig. 4. This technique is based on gradually widening sections of a line to decrease its impedance and hence encouraging the signal to move forward and not reflecting back in the wrong direction. The idea can be used in both gate and drain lines. In order to design a tapered DA, the first thing to consider is that the $Z_0$ is not a single constant of 50 Ω. The characteristic impedance of the lines is $Z_0$=50 Ω only at input and output ports in order to fulfill the matching. We can reduce the impedances smoothly along the lines but in order to make the calculations easier, each stage can get estimated with a specific characteristic impedance. Therefore, there will be discontinuities between stages. Now we will use the theory of small reflections to compute the total reflection coefficient [2]. According to [2], the overall reflection coefficient of a tapered design is:

$$\Gamma_{overall}(\theta) = \Gamma_0 + \Gamma_1 e^{-2j\theta} + \Gamma_2 e^{-4j\theta} $$
$$+ ... + \Gamma_n e^{-2nj\theta} \qquad (29)$$

where $\Gamma_i$ is reflection coefficient between stage i and its next stage i+1, n is the number of the stages, and θ is the electrical length of the sections in the center frequency of the bandwidth (equal-length sections). To choose a proper amount for θ, two points must be considered. First, in order to have a real number as the overall reflection coefficient, θ should be a multiple of π/2. Second, if θ is an even multiple of π/2, then the partial reflection coefficients get added to each other which means lots of reflections and losing the signals. Hence, we should choose θ to be an odd multiple of π/2. In this situation, the sign of each part will alter in a way that we can reach to a small overall reflection coefficient. That is a worthwhile result. Assuming θ= π/2, using (29) the resulting $\Gamma_{overall}$ will be:

$$\Gamma_{overall}(\theta) = \Gamma_0 + \Gamma_1(e^{-j\pi})^1 + \Gamma_2(e^{-j\pi})^2$$
$$+...+\Gamma_n(e^{-j\pi})^n = \Gamma_0 + \Gamma_1(-1)^1 + \Gamma_2(-1)^2 \qquad (30)$$
$$+...+\Gamma_n(-1)^n$$

$$\Gamma_{overall}(\theta) = \Gamma_0 - \Gamma_1 + \Gamma_2 - ... + \Gamma_n(-1)^n \qquad (31)$$

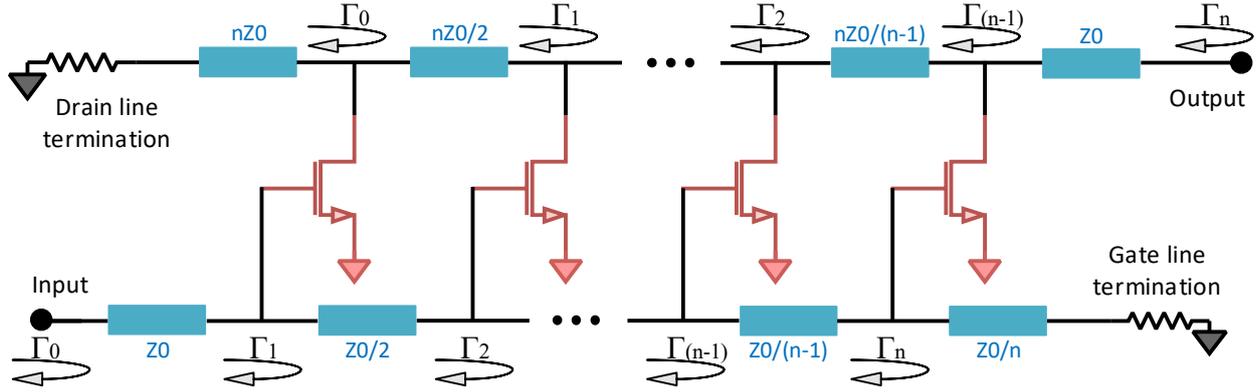

**Fig. 4** A DA design using tapering method in gate and drain lines.

Now, we are able to evaluate the equivalent characteristic impedance of the gate and drain lines.

$$\frac{Z_{overall(G)} - (50\Omega)}{Z_{overall(G)} + (50\Omega)} = \Gamma_{overall(G)} \qquad (32)$$

$$\rightarrow Z_{overall(G)} = (50\Omega)(\frac{1 + \Gamma_{overall(G)}}{1 - \Gamma_{overall(G)}})$$

$$\frac{(50\Omega) - Z_{overall(D)}}{(50\Omega) + Z_{overall(D)}} = \Gamma_{overall(D)} \qquad (33)$$

$$\rightarrow Z_{overall(D)} = (50\Omega)(\frac{1 - \Gamma_{overall(D)}}{1 + \Gamma_{overall(D)}})$$

As a result, the cut-off frequency of the gate and drain lines can get estimated using (4) and (5).

$$f_{cut-off(G)} = \frac{1}{\pi Z_{overall(G)} C_{GS}} \qquad (34)$$

$$f_{cut-off(D)} = \frac{1}{\pi Z_{overall(D)} C_{DS}} \qquad (35)$$

Consequently, the frequency limit of the design will be set by the smaller value of (34) or (35). In this step, to have a better understanding of the procedure, let us apply it to a known tapering pattern which is first used in [12]. In this approach, along the gate line, $Z_{0(G)}$ declines from 50 Ω to $Z_0/n$, and along the drain line, $Z_{0(D)}$ decreases from $nZ_0$ to reach 50 Ω at the output port. Each section of the lines will have a different cut-off frequency if left on their own. The smallest cut-off frequency of these sections could be the cut-off frequency of the whole circuit, if these cut-off frequencies were sufficiently far apart and we could neglect the reflections which happen at the discontinuities. As these two conditions are not fulfilled, we employ (31) once again to approximate the overall reflection coefficients of the gate and drain lines in the tapered approach, shown in Fig. 4. For gate line:

$$\Gamma_{overall(G)}(\theta) = \Gamma_0 - \Gamma_1 + \Gamma_2 - ... + \Gamma_n(-1)^n = 0 = \frac{\frac{Z_0}{2} - Z_0}{\frac{Z_0}{2} + Z_0} + \frac{\frac{Z_0}{3} - \frac{Z_0}{2}}{\frac{Z_0}{3} + \frac{Z_0}{2}}$$

$$\frac{\frac{Z_0}{4} - \frac{Z_0}{3}}{\frac{Z_0}{4} + \frac{Z_0}{3}} + \frac{\frac{Z_0}{5} - \frac{Z_0}{4}}{\frac{Z_0}{5} + \frac{Z_0}{4}} - ... + \frac{\frac{Z_0}{n} - \frac{Z_0}{(n-1)}}{\frac{Z_0}{n} + \frac{Z_0}{(n-1)}}(-1)^n \qquad (36)$$

And for drain line it will be:

$$\Gamma_{overall(D)}(\theta) = \Gamma_0 - \Gamma_1 + \Gamma_2 - ... + \Gamma_n(-1)^n = 0 = \frac{\frac{nZ_0}{(n-1)} - Z_0}{\frac{nZ_0}{(n-1)} + Z_0}$$

$$\frac{\frac{nZ_0}{(n-2)} - \frac{nZ_0}{(n-1)}}{\frac{nZ_0}{(n-2)} + \frac{nZ_0}{(n-1)}} - ... + \frac{nZ_0 - \frac{nZ_0}{2}}{nZ_0 + \frac{nZ_0}{2}}(-1)^n \qquad (37)$$

So, in the case of tapering technique (36) and (37) will give the reflection coefficient of the gate and the drain line respectively. By putting their answers into (32) and (33), $Z_{overall(G)}$ and $Z_{overall(D)}$ will be achieved. Then by using (34) and (35), the cut-off frequency of gate and drain lines are known. And the smaller value of these two will limit the overall frequency behavior and will be the maximum achievable frequency. The value in that condition is lower than the value achieved by equation (4). As an example, let's look at the design in [15]. It has 4 stages and it has used the tapering technique. Considering (36) and (37), for $Z_0$=50 Ω and n=4, the overall reflection coefficients of the gate and the drain lines are:

$$\Gamma_{overall\,(G)}(\theta) = 0 + 0.33 - 0.2 + 0.14 - 0.11 = 0.16 \qquad (38)$$

$$\Gamma_{overall\,(D)}(\theta) = -0.33 + 0.2 - 0.14 + 0 = -0.27 \qquad (39)$$

Therefore, the equivalent characteristic impedances of the gate and drain lines are calculated using (32) and (33), respectively.

$$Z_{overall\,(G)} = (50\Omega)(\frac{1 + \Gamma_{overall\,(G)}}{1 - \Gamma_{overall\,(G)}}) = 69\ \Omega \qquad (40)$$

$$Z_{overall\,(D)} = (50\Omega)(\frac{1 - \Gamma_{overall\,(D)}}{1 + \Gamma_{overall\,(D)}}) = 87\ \Omega \qquad (41)$$

The work presented in [15] uses TGF2023-01 transistor from Qorvo and the $C_{GS}$ is 1.79 pF. So, the cut-off frequency of the lines using (34) and (35) will be:

$$f_{cut-off\,(G)} = \frac{1}{\pi Z_{overall\,(G)} C_{GS}} = 2.58\ \text{GHz} \qquad (42)$$

$$f_{cut-off\,(D)} = \frac{1}{\pi Z_{overall\,(D)} C_{DS}} = 11.87\ \text{GHz} \qquad (43)$$

And the resulting cut-off frequency of the whole circuit is:

$$\rightarrow f_{cut-off\,(total)} = \min(f_{cut-off\,(G)}, f_{cut-off\,(D)}) \qquad (44)$$
$$= 2.58\ \text{GHz}$$

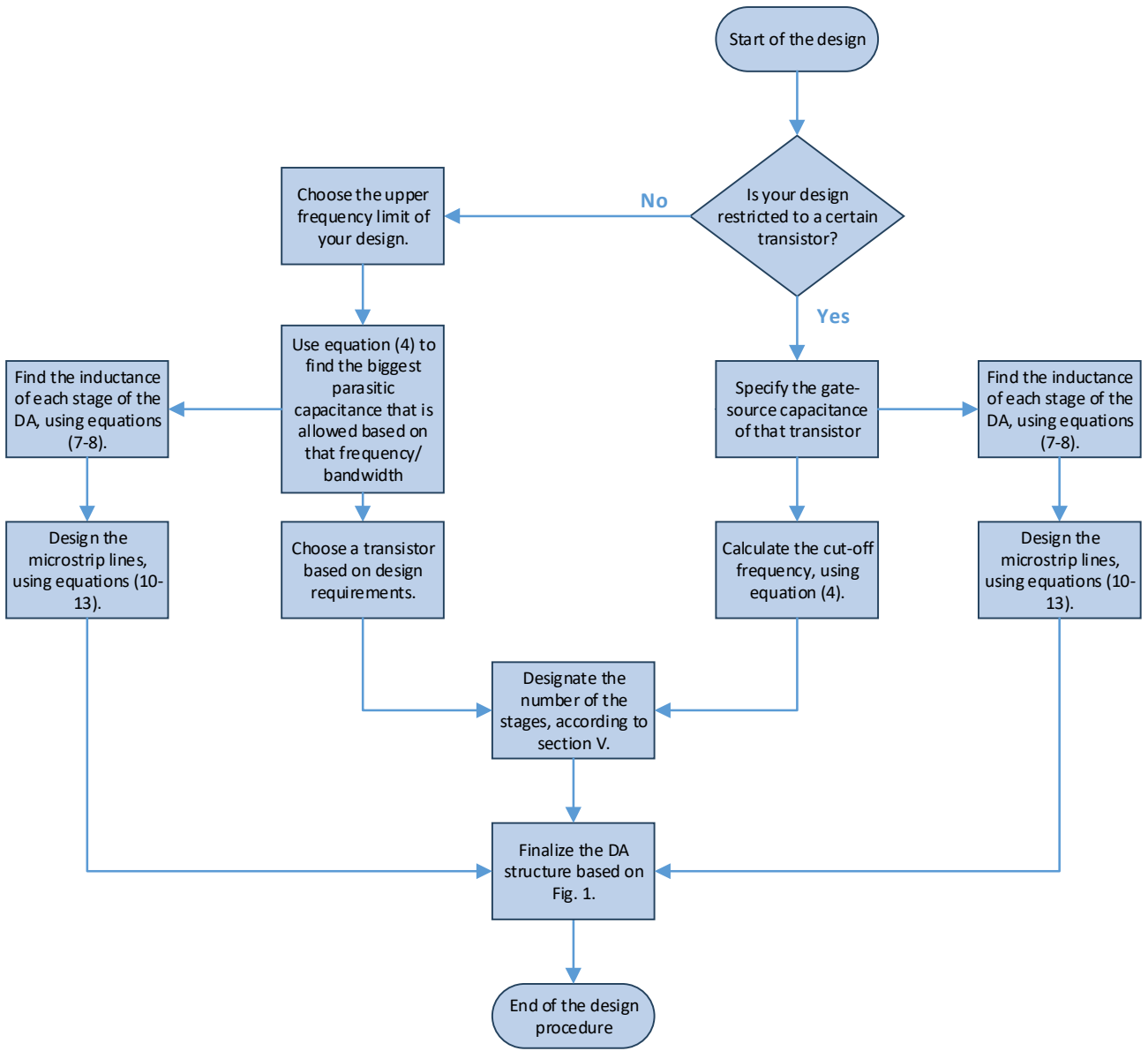

**Fig. 5** The suggested design procedure.

We can see that there is an important trade-off when designing a tapered DA. If we choose a very big $Z_0$ to start from, the efficiency will be boosted and the reflected signals diminish drastically. However, we will have a very small bandwidth. And if we design the tapered DA for a higher bandwidth, the efficiency will not improve noticeably. Therefore, we should find an optimum point to achieve the desired bandwidth and efficiency.

## 6. Design Procedure

The corresponding flowchart of the discussed design procedure is demonstrated in Fig. 5. The flowchart is summarizing the paths which are in front of a designer while initiating the design of a distributed amplifier. In the start, a designer will have either a targeted bandwidth or a specific transistor type to begin with. According to Fig. 5, the discussions in this paper, can help the former case to choose a transistor for their target and can help the latter case to know what frequency behavior to expect. Let's now, by the help of some related published works, take a look at how well the discussed design procedure works. Those published works are mentioned in Table I. In published works, circuits are fabricated, final measurements are done and transistors are already chosen, so we will take the left side of the flowchart. As the first case, let's return to the example from section V, [15] made use of TGF2023-01 transistor from Qorvo with $C_{GS}$=1.79 pF. The flowchart refers us to (4). According to (4) the best cut-off frequency of a conventional DA (i.e. without manipulating $C_{GS}$) can be 3.55 GHz when using that particular transistor. In order to improve the power gain and efficiency, [15] has also used tapering technique in both of the gate and the drain lines. As we expect from section V, the use of tapering technique limited the possible bandwidth to 2.58 GHz. To solve that, [15] compensated the bandwidth reduction with inserting an input matching network. As a result, the bandwidth approached the maximum possible frequency limit, i.e. 3.55 GHz. Therefore, we are seeing that the performance enhancement came at a price of a complicated design and more importantly a bandwidth reduction. The other example of a published work is [16]. It used 0.18 μm GaN HEMT process and the biggest transistors have 200 μm gate periphery. So their $C_{GS}$ is approximately $C_{ox}*W*L$=124 fF; and (4) predicts that such design could reach to 51 GHz at maximum. Again tapering technique is used and a reduction in the bandwidth would happen. However, the design has benefitted from some gate series capacitances that lowered the effective capacitance in the gates and compensated for some of the consequences of tapering technique. The design in [17] used SMIC 40 nm CMOS and all the transistors are 100 μm wide, so the $C_{GS}$ is estimated to be $C_{ox}*W*L$=138 fF. The answer which (4) gives is 46 GHz, but the maximum frequency in the measured bandwidth is 34 GHz and it is due to the uncompensated capacitances at the internal nodes of the stacked cascode cells, consisting of three transistors. However, using cascode structure in cells of a distributed amplifier has several advantages too, such as higher power gain and better isolation between input and output which leads to a better stability. Similar discussions can be made upon the other works, but here in order to be concise we do not go into much details of the other rows of Table I. In [4], capacitors are placed in series with the gates of the transistors. This technique has improved the bandwidth significantly. But according to Table I it could be even higher if double-stacked transistors or cascode construction was not implemented. For the reason that there is a capacitance at the cascode node that does not get absorbed in the transmission lines of the DA structure. So it limits the bandwidth. The same constraint is true for [9] as there is an additional capacitance at the cascode node which limits the bandwidth. In [14], this parasitic capacitance is mitigated by the help of a series inductor at cascode node, but probably its other design parameters prevent the bandwidth from reaching the upper frequency limit.

**Table 1**    Verification of our approach using some recent works.

| Ref. | $P_{out}$ (watts) | $PAE_{max}$ (%) | $Gain_{max}$ (dB) | Achieved Bandwidth (GHz) | Effective Capacitance | Estimated upper frequency limit based on (4) and (5) |
|------|------|------|------|------|------|------|
| [4] | 0.06 | 12.5 | 10.5 | 1-160 | 20 fF | 318 GHz |
| [5] | 0.12 | 14.5 | 18.3 | 12-46 | 97 fF | 65.6 GHz |
| [9] | 0.02 | 5-22 | 7-13 | 1-10 | 0.28 pF | 22.7 GHz |
| [10] | 10.6-24.3 | 15.5-26.6 | 15.3-23.2 | 6-18 | 0.3 pF | 21.2 GHz |
| [13] | 14.5-26.3 | 13.2-23.7 | 17-21 | 6-18 | 0.3 pF | 21.2 GHz |
| [14] | 0.015 | 4-10.6 | 9.8 | 1-15.2 | 0.14 pF | 45 GHz |
| [15] | 13 | 27 | 22 | DC-3.4 | 1.79 pF | 3.55 GHz |
| [16] | 1.26-2.19 | 9.4-16.8 | 3-5.5 | 5-38 | 124 fF | 51 GHz |
| [17] | 0.088 | 6 | 22 | 14-34 | 138 fF | 46 GHz |

## 7. Conclusion

In this paper, we proposed a design procedure for a DA and presented an analytical estimation of the bandwidth. The upper frequency limit of a DA structure is the cut-off frequency of the most limiting section of its artificial transmission lines. We discussed the maximum achievable bandwidth of a conventional DA with and without using series capacitors at the gates of the transistors. Furthermore, this paper discusses the frequency behavior of a tapered DA for the first time and provides simple analytical equations for predicting its gate and drain cut-off frequencies. The procedure and the discussions in this paper provide fast, straightforward and accurate predictions for designer of any types of DA.

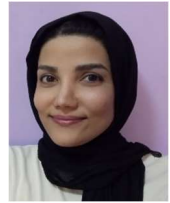

**Elham Amiri** received the B.Sc. degree in electrical engineering from the AmirKabir University of Technology, Tehran, Iran, in April 2010, the M.Sc. degree (Hons.) in electrical engineering from the Ferdowsi University of Mashhad, Mashhad, Iran, in Sep. 2014. She is currently pursuing the Ph.D. degree in electrical engineering at Ferdowsi University of Mashhad, Mashhad, Iran. In 2021, she was a visiting research student at University of Alberta, Edmonton, Canada. Her research interests are high frequency circuits, integrated circuit design, mixed signal design, sensors and energy harvesting.

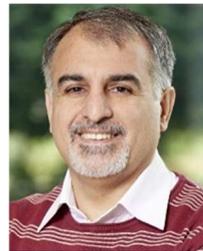

**Mojtaba Joodaki** received the B.Sc. degree in Electrical and Electronic engineering from Iran Science and Technology University, Tehran, Iran, in 1994, and the M.Sc. degree in Electrical and Electronic engineering from Tarbiat Modarres University, Tehran, in 1997, and the Ph.D. degree in Electrical Engineering from Kassel University, Kassel, Germany, in 2002. He joined ATMEL Germany GmbH, Heilbronn, Germany, as a Device Engineer in 2002, where he was working on technology development, modeling, and characterization of Si- and SiGe-based devices for RF applications. In April 2005, he joined Infineon Technologies AG, Munich, Germany, as a Development Engineer, where he was responsible for electromagnetic interference (EMI)/electromagnetic compatibility (EMC) of memory modules. In his last industrial position, from October 2006 to July 2009, he was a Device Engineer with Qimonda GmbH, Dresden, Germany, involved in developing nanotransistors for dynamic random-access memory (DRAM) products. Then, he started working as a Visiting Scientist and a Lecturer with the Institute of Nanostructure Technologies and Analytics (INA), Kassel University, where he defended his Habilitation dissertation in April 2011. In 2010, he joined the Department of Electrical Engineering, Ferdowsi University of Mashhad, Mashhad, Iran, as an Assistant Professor of Electronic Engineering (RF Circuit Design and Semiconductor Devices and Technology), where he was promoted to Associate Professor in September 2011 and a Professor in July 2018. Since January 2020, he has been with Constructor University (former Jacobs University), Bremen, Germany, as a Professor of Electrical Engineering and Electrical, and the Computer Engineering Program Chair. His research interests include modeling, characterization, and fabrication of passive and active devices (organic and inorganic) for high-frequency, memory, and optoelectronic applications as well as EMC of electronic products. Mr. Joodaki is a Life Member of the International Society for Optics and Photonics (SPIE). He has been awarded several prizes for his scientific activities, including the Best Dissertation Prize of North Hessen Universities from the Association of German Engineers (VDI) in 2004, the Young Graduated Research Fellowship of the Gallium Arsenide (GaAs) Association in the European Microwave Week 2001 and 2002, and the F-Made Scholarship of SPIE 2002.